\documentclass[epj]{webofc}
\usepackage[varg]{txfonts}
\usepackage{amsmath}
\usepackage{amsfonts}
\usepackage{amssymb}
\usepackage{graphicx}
\usepackage[toc,page]{appendix}

\wocname{EPJ Web of Conferences}
\woctitle{ICNFP 2017}
\newcommand{\lp}{\left(}
\newcommand{\rp}{\right)}
\newcommand{\lsb}{\left[}
\newcommand{\rsb}{\right]}
\newcommand{\bea}{\begin{eqnarray}}
\newcommand{\eea}{\end{eqnarray}}
\newcommand{\beal}[1]{\begin{eqnarray}\label{#1}}
\newcommand{\eeal}{\end{eqnarray}}
\newcommand{\nn}{\nonumber}
\newcommand{\f}[2]{\frac{#1}{#2}}
\newcommand{\EQ}[1]{Eq.~(\ref{#1})}
\newcommand{\EQS}[1]{Eqs.~(\ref{#1})}

\newcommand{\rfn}[1]{(\ref{#1})}

\newcommand{\p}{\partial}
\newcommand{\twpt}{(\tau,w,\pT)}

\newcommand{\s}{{\rm s}}	% species indexing
\newcommand{\Qp}{{Q^+}}		% species indexing
\newcommand{\Qm}{{Q^-}}		% species indexing
\newcommand{\G}{{G}}		% species indexing
\newcommand{\Qpm}{{Q^\pm}}	% species indexing
\newcommand{\Q}{{Q}}		% species indexing
\newcommand{\eq}{{\rm eq}}  %  equilibrium
\newcommand{\an}{{\rm a}} 	%  anisotropic
\newcommand{\teq}{\tau_\eq}
\newcommand{\hI}{\hat{{\cal I}}}
\newcommand{\pT}{p_{T}}

\def\Lq{\Lambda_Q}

\def\Lg{\Lambda_G}

\def\xq{\xi_Q}
\def\xg{\xi_G}

\newcommand{\VP}{\vphantom{\frac{}{}}\!}
\def\l{\limits}
\begin{document}
\title{Kinetic equations and anisotropic hydrodynamics \\ for quark and gluon fluids}
\author{\firstname{Ewa} \lastname{Maksymiuk}\inst{1}\fnsep\thanks{\email{MaksymiukEwa@gmail.com}} 
}

\institute{Institute of Physics, Jan Kochanowski University, PL-25-406~Kielce, Poland
          }

\abstract{%
The mixture of quark and gluon fluids is studied in a one-dimensional boost-invariant setup using the set of relativistic kinetic equations treated in the relaxation time approximation. Effects of a finite quark mass, non-zero baryon number density, and quantum statistics are discussed. Comparisons between the exact kinetic-theory results and anisotropic hydrodynamics predictions are performed and a very good agreement between the two are found.
}
\maketitle
\section{Introduction}
\label{intro}
Despite a significant development of theoretical tools and experimental methods in the last years, an unambiguous interpretation of data measured in the heavy-ion collision experiments at RHIC and the LHC is still challenging~\cite{Florkowski:2010zz}. One of the basic and most successful tools used for this purpose is relativistic viscous hydrodynamics~\cite{Ollitrault:2012cm,Gale:2013da,Jeon:2015dfa,Jaiswal:2016hex,Florkowski:2017olj}. Unfortunately, it turns out that its analytic structure and applicability range are not completely settled yet~\cite{Florkowski:2017olj}. Recently, new methods have been proposed which allow, to some extent, to test the validity of different hydrodynamic formalisms~\cite{Florkowski:2013lza,Florkowski:2013lya,Florkowski:2014sfa,Denicol:2014tha,Denicol:2014xca,Florkowski:2014sda}. These methods are  based on direct comparisons between predictions of various hydrodynamic models and exact solutions of the underlying kinetic-theory equations ~\cite{Florkowski:2013lza,Florkowski:2013lya,Florkowski:2014sfa,Denicol:2014tha,Denicol:2014xca,Florkowski:2014sda,Bazow:2013ifa,Nopoush:2014qba,Denicol:2014mca,Jaiswal:2014isa,Chattopadhyay:2014lya,Florkowski:2015lra, Tinti:2015xra,Molnar:2016gwq,Martinez:2017ibh}. In particular, they have been applied to study mixtures of quark and gluon fluids~\cite{Florkowski:2012as,Florkowski:2013uqa,Florkowski:2014txa,Florkowski:2015cba}.

In this proceedings contribution we continue studies of the quark-gluon mixture within the kinetic-theory setup  and generalize it by including the finite mass of quarks, non-zero baryon number density, and quantum statistics effects \cite{Florkowski:2017jnz}. Moreover, following ideas of the anisotropic hydrodynamics (aHydro) \cite{Florkowski:2010cf,Martinez:2010sc}, we formulate the evolution equations devoted to describe such a mixture of fluids~\cite{Florkowski:2017ovw}. Since the aHydro formalism is based on the idea that the high pressure anisotropy of the produced matter may be included already in the leading order of the hydrodynamic expansion, our treatment may be especially useful to describe early stages of the quark-gluon plasma evolution in the ultra-relativistic regime of heavy-ion collisions.
%
%%%%%%%%%%%%%%%%%%%%%%%%%%%%%%%%%%%%%%%%%%%%%%%%%%
\section{Kinetic equations}
\label{Kin_eq}
%%%%%%%%%%%%%%%%%%%%%%%%%%%%%%%%%%%%%%%%%%%%%%%%%%
%
We start our analysis with coupled kinetic equations for quarks, antiquarks, and gluons, 
\beal{kineq}
\lp p\cdot\p \rp  f_{\s}(x,p) \!\!\!\!&=&\!\!\!\!  {\cal C}\lsb  f_{\s}(x,p)\rsb,\quad \s=\Qp, \Qm,G. 
\eeal
For simplicity, we treat the collisional kernel ${\cal C}$ in Eqs.~(\ref{kineq})  in the relaxation time approximation~(RTA)~\cite{Bhatnagar:1954zz,Anderson:1974a,Anderson:1974b,Czyz:1986mr}, namely
\beal{colker}
{\cal C}\lsb f_\s(x,p)\rsb \!\!\!\! &=&\!\!\!\!   \lp p\cdot U \rp \frac{f_{\s, \eq}(x,p)-f_\s(x,p)}{\teq},
\eeal
where $U(x)$ is the four-velocity of the fluid element and $\teq$ is the relaxation time, which in this work is chosen to be constant.
%
%%%%%%%%%%%%%%%%%%%%%%%%%%%%%%%%%%%%%%%%%%%%%%%%%% 
\subsection{Equilibrium distributions}
\label{ssec:ed}
%%%%%%%%%%%%%%%%%%%%%%%%%%%%%%%%%%%%%%%%%%%%%%%%%% 
%
The equilibrium distribution functions, $f_{\s, \eq}(x,p)$, introduced in \EQ{colker} have the Fermi-Dirac or Bose-Einstein forms, 
\beal{feq}
f_{\Qpm, \eq}(x,p) = h^{+}_\eq\lp\f{  p\cdot U  \mp \mu }{T }\rp ,
\quad
f_{\G, \eq}(x,p) = h^{-}_\eq\lp\f{p\cdot U}{T }\rp,
\eeal
for (anti)quarks and gluons, respectively. Here we define
\beal{heq}
h^{\pm}_\eq(a)\!\!\!\! &=& \!\!\!\!  \lsb\VP\exp(a) \pm 1 \rsb^{-1},
\eeal
with  ,,$+$'' corresponding to fermions and ,,$-$''  describing bosons. In Eq.~(\ref{colker}) it is assumed that all considered particle species approach the same local equilibrium state characterized by the effective temperature $T(x)$ and effective baryon chemical potential $\mu(x)$. 
%
%%%%%%%%%%%%%%%%%%%%%%%%%%%%%%%%%%%%%%%%%%%%%%%%%%
\subsection{Formal solutions}
\label{ssec:boost-inv-eq}
%%%%%%%%%%%%%%%%%%%%%%%%%%%%%%%%%%%%%%%%%%%%%%%%%%
%
In general, due to the complex structure of Eqs.~(\ref{kineq}), it is not possible to solve them exactly. For this reason we assume the Bjorken flow symmetry \cite{Bjorken:1982qr}, which states that the system is boost-invariant along the beam ($z$) direction and homogeneous in the transverse ($xy$) plane. The Bjorken symmetry is a reasonable approximation for the midrapidity region at the LHC energies. In this case, the distribution functions depend solely on the longitudinal proper time $\tau=\sqrt{t^2-z^2}$, the transverse momentum \mbox{$\pT = \sqrt{p_x^2+p_y^2}$}, and the boost-invariant variable $w=t p_L -z E_p$ \cite{Czyz:1986mr}. These assumptions allow us to find exact formal solutions of Eqs.~(\ref{kineq}) which have the following simple form~\cite{Florkowski:2013lza,Florkowski:2013lya,Baym:1984np,Baym:1985tna} 
\beal{formsolQ}
f_\s\twpt = D(\tau,\tau_0) f_\s^0(w,\pT) + \int\l_{\tau_0}^{\tau} \f{d \tau'}{\teq^\prime}\  D(\tau,\tau') 
f_{\s,\eq}(\tau^\prime,w,p_T).
 \eeal
Here $f_{\s,\eq}(w,p_T)$ are the equilibrium distribution functions which may be obtained from \EQ{feq}  using relations $p\cdot U=v/\tau$ and $v=\sqrt{w^2+(m^2+\pT^2)\tau^2}$, while 
$D(\tau_2,\tau_1)= \exp\lp(\tau_1-\tau_2)\tau_\eq\rp$ is the so-called damping function.
Solutions (\ref{formsolQ}) are used to find the form of all thermodynamic-like variables characterizing the system (such as particle density, ${\cal N}$, energy density, ${\cal E}$, transverse, ${\cal P}_T$, and longitudinal, ${\cal P}_L$, pressures) using the canonical kinetic theory definitions of these quantities.
%
%%%%%%%%%%%%%%%%%%%%%%%%%%%%%%%%%%%%%%%%%%%%%%%%%%
\subsection{Anisotropic distributions}
\label{ssec:ad}
%%%%%%%%%%%%%%%%%%%%%%%%%%%%%%%%%%%%%%%%%%%%%%%%%%
%
In order to study the evolution of the system defined by \EQS{kineq} towards local thermal equilibrium we allow it to be initially locally anisotropic in the momentum space. This may be achieved by assuming that the initial distributions are given by the anisotropic Romatschke-Strickland (RS) forms~\cite{Romatschke:2003ms}. In the covariant versions they read~\cite{Florkowski:2012as}

\beal{fa}\nn
f_{\Qpm, \an}(x,p) \!\!\!\!&=&\!\!\!\! h^{+}_\eq\lp\f{\sqrt{\lp p \cdot U\rp^2+\xi_\Q  \lp p \cdot Z\rp^2} \mp \lambda }{\Lambda_\Q}\rp,
\\ 
f_{\G, \an}(x,p) \!\!\!\!&=&\!\!\!\! h^{-}_\eq\lp\f{\sqrt{\lp p \cdot U\rp^2+\xi_\G  \lp p \cdot Z\rp^2}}{\Lambda_G}\rp.\ 
\eeal
The anisotropy parameter, $\xi_\Q(x)$, and the transverse-momentum scale, $\Lambda_\Q(x)$, in \EQS{fa} are the same for quarks and antiquarks. 
The function $\lambda(x)$ is the non-equilibrium baryon chemical potential of quarks and antiquarks. Similarly, $\xi_G(x)$ is the gluon anisotropy parameter and $\Lambda_G(x)$ is the gluon transverse-momentum scale.
%

%
%%%%%%%%%%%%%%%%%%%%%%%%%%%%%%%%%%%%%%%%%%%%%%%%%% 
\subsection{Initial distributions}
\label{ssec:id}
%%%%%%%%%%%%%%%%%%%%%%%%%%%%%%%%%%%%%%%%%%%%%%%%%% 
% 
Using boost-invariant variables $w$ and $v$, and the relation $p \cdot Z = -w/\tau$ in \EQS{fa}  we may express the initial distribution functions is \EQ{formsolQ} as follows
\beal{f0}
&& f_{\Qpm}^0 (w,p_T) = 
h^{+}_\eq \lp\f{\sqrt{\lp1+\xi_\Q^0\rp\lp\frac{w}{\tau_0}\rp^2+  m^2+\pT^2   } \mp \lambda ^0 }{\Lambda_\Q^0}   \rp  ,
\\
&& f_{\G}^0 (w,p_T) = h^{-}_\eq\lp\f{\sqrt{\lp1+\xi_\G^0\rp\lp\frac{w}{\tau_0}\rp^2+\pT^2   }}{\Lambda_\G^0} \,  \rp,
\eeal
where $f_\s^0(w,p_T)\equiv f_\s (\tau_0,w,p_T)$. Here we introduced initial parameters $\xi_\s^0\equiv\xi_\s(\tau_0)$, $\Lambda_\s^0\equiv\Lambda_\s(\tau_0)$,  and $\lambda^0\equiv\lambda(\tau_0)$ with $\tau_0$ being the initial time for the system evolution.
%%%%%%%%%%%%%%%%%%%%%%%%%%%%%%%%%%%%%%%%%%%%%%%%%%
\subsection{Baryon number and four-momentum conservation laws}
\label{ssec:LM}
%%%%%%%%%%%%%%%%%%%%%%%%%%%%%%%%%%%%%%%%%%%%%%%%%%
%
The effective temperature, $T(x)$, and effective baryon chemical potential, $\mu(x)$, are found by assuming that baryon number and four-momentum are conserved during the space-time evolution, which results in the so-called Landau matching conditions,
\beal{LMs}
{\cal B}^\eq = {\cal B},
\quad
{\cal E}^\eq = {\cal E},
\eeal
where the baryon number density  is ${\cal B} =  \lp {\cal N}_{\Qp}-{\cal N}_{\Qm} \rp/3 $. Equations~(\ref{LMs}) may be explicitly written as follows
\beal{FIRST-EQUATION}
T^3 \sinh\lp\frac{\mu}{T}\rp\,{\cal H}_{\cal B}\lp \frac{m}{T},   \frac{\mu}{T}\rp  
\!\!\!\!&=&\!\!\!\!
   \f{\tau_0 \lp\Lambda_\Q^0\rp^3}{\tau \sqrt{1+\xi_\Q^0}} \sinh\lp\frac{\lambda^0}{\Lambda_\Q^0}\rp\,{\cal H}_{\cal B}\lp \frac{m}{\Lambda_\Q^0},\frac{\lambda^0}{\Lambda_\Q^0} \rp D(\tau,\tau_0) \nn \\
&+&\!\!\! \int\l_{\tau_0}^{\tau} \f{d \tau'}{\teq^\prime}\  D(\tau,\tau')  \f{\tau^\prime \lp T^\prime\rp^3}{\tau  } \sinh\lp\frac{\mu^\prime}{T^\prime}\rp\,{\cal H}_{\cal B}\lp \frac{m}{T^\prime}, \frac{\mu^\prime}{T^\prime}\rp
\eeal
and 
\beal{SECOND-EQUATION}
 &&T^4 \lsb \tilde{{\cal H}}^+\lp 1, \frac{m}{T}, - \frac{\mu}{T}\rp +
 \tilde{{\cal H}}^+\lp 1, \frac{m}{T}, + \frac{\mu}{T}\rp  + r  \tilde{{\cal H}}^-\lp 1, 0,0\rp \rsb \nn \\
 &=&     \lp\Lambda_\Q^0\rp^4 
\lsb \tilde{{\cal H}}^+\lp \f{\tau_0}{\tau \sqrt{1+\xi_\Q^0}}, \frac{m}{\Lambda_\Q^0}, - \frac{\lambda^0}{\Lambda_\Q^0}\rp 
+ \tilde{{\cal H}}^+\lp \f{\tau_0}{\tau \sqrt{1+\xi_\Q^0}}, \frac{m}{\Lambda_\Q^0}, + \frac{\lambda^0}{\Lambda_\Q^0}\rp \rsb
D(\tau,\tau_0) \nn \\
&&+    \int\l_{\tau_0}^{\tau} \f{d \tau'}{\teq^\prime}\  D(\tau,\tau') \lp T^\prime\rp^4
\lsb \tilde{{\cal H}}_{\cal  }^+\lp \f{\tau^\prime}{\tau  }, \frac{m}{T^\prime}, - \frac{\mu^\prime}{T^\prime}\rp  +
\tilde{{\cal H}}_{\cal  }^+\lp \f{\tau^\prime}{\tau  }, \frac{m}{T^\prime}, + \frac{\mu^\prime}{T^\prime}\rp   \rsb 
 \\
 &&+  r  \lsb  \lp\Lambda_\G^0\rp^4 \tilde{{\cal H}}^-\lp \f{\tau_0}{\tau \sqrt{1+\xi_\Q^0}},  0,0\rp D(\tau,\tau_0) 
+   \int\l_{\tau_0}^{\tau} \f{d \tau'}{\teq^\prime}\  D(\tau,\tau') \lp T^\prime\rp^4 \tilde{{\cal H}}_{\cal  }^-\lp \f{\tau^\prime}{\tau  }, 0,0\rp\rsb, \nn 
\eeal
respectively.
Functions  $\tilde{{\cal H}}^\pm$ and ${\cal H}_{\cal B}$  are defined in Ref.~\cite{Florkowski:2017jnz}. 
%
%%%%%%%%%%%%%%%%%%%%%%%%%%%%%%%%%%%%%%%%%%%%%%%%%%
%%%%%%%%%%%%%%%%%%%%%%%%%%%%%%%%%%%%%%%%%%%%%%%%%%
%%%%%%%%%%%%%%%%%%%%%%%%%%%%%%%%%%%%%%%%%%%%%%%%%%
\section{Anisotropic hydrodynamics}
\label{sect:aHydro}
%%%%%%%%%%%%%%%%%%%%%%%%%%%%%%%%%%%%%%%%%%%%%%%%%%
%%%%%%%%%%%%%%%%%%%%%%%%%%%%%%%%%%%%%%%%%%%%%%%%%%
%%%%%%%%%%%%%%%%%%%%%%%%%%%%%%%%%%%%%%%%%%%%%%%%%%
%
In this section we introduce the aHydro scheme in a version appropriate for mixtures of fluids. This framework is expected to describe well soft modes of systems exhibiting large pressure anisotropies. The aHydro formalism is build upon the assumption that at leading order the studied system is anisotropic in the momentum space and during the \emph{entire} space-time evolution may be well approximated by the RS ansatz (\ref{fa}). The equations of motion in such a case may be most easily  obtained by taking moments of the kinetic equations~(\ref{kineq}) with the  RS distribution functions treated as solutions (i.e., with the substitution $f_{\s}(x,p)\to f_{\s, \an}(x,p)$) and by assuming elementary conservation laws. 

To find moments of the kinetic equations~(\ref{kineq}) it is very convenient to introduce the $n$-th moment integral operator in the momentum space 
\beal{Ioperator}
\hI^{\mu_1\cdots\mu_n} (\dots) \equiv  \int \!dP\, p^{\mu_1}p^{\mu_2}\cdots p^{\mu_n} (\dots).
\eeal
Then the $n$-th moments of the kinetic equations~(\ref{kineq}) are found by acting with $\hI^{\mu_1\cdots\mu_n}$  on their left- and right-hand sides and multiplying them by the degeneracy factors $k_s$, with $k_\Qpm=12/(2\pi)^3$ and $k_G=16/(2\pi)^3$, namely
\beal{nth_mom}
k_\s \hI^{\mu_1\cdots\mu_n}  p^\mu \p_\mu  f_\s(x,p) 
\!\!\!\! &=& \!\!\!\!
 k_\s \hI^{\mu_1\cdots\mu_n} p^\mu U_\mu\frac{f_{\s, \eq}(x,p)-f_\s(x,p)}{\teq}.
\eeal
In the following sections we show  how to use the zeroth, first, and second moments of the kinetic equations~(\ref{kineq}) to obtain the closed system of aHydro evolution equations. The latter allow us to determine seven unknown functions: $T(\tau)$, $\mu(\tau)$, $\xi_Q(\tau)$, $\xi_G(\tau)$, $\Lq(\tau)$, $\Lg(\tau)$, and $\lambda(\tau)$, which fully define distributions (\ref{f0}). These are necessary to obtain physical observables, in particular,  ${\cal E}^\an$, ${\cal P}_T^\an$, and ${\cal P}_L^\an$.
%
%%%%%%%%%%%%%%%%%%%%%%%%%%%%%%%%%%%%%%%%%%%%%%%%%% 
\subsection{Zeroth moments of the kinetic equations}
\label{sect:0th_m}
%%%%%%%%%%%%%%%%%%%%%%%%%%%%%%%%%%%%%%%%%%%%%%%%%%
%
The zeroth moments of the kinetic equations \rfn{kineq} give three scalar equations
\beal{0mq}
\p_\mu  \left( {\cal N}_{s, \rm a} U^\mu \right)   &=& \frac{{\cal N}_{s, \rm eq} -{\cal N}_{s, \rm a} }{\teq}.
\eeal
Subtracting the antiquark component of \EQS{kineq} from the quark one   and making use of the definition of the baryon density leads to the following constraint
\beal{Ba1}
\f{d{\cal B}_{\rm a}(\tau)}{d\tau} +\f{{\cal B}_{\rm a}(\tau)}{\tau} = \f{{\cal B}_{\rm eq} - {\cal B}_{\rm a}}{\teq}.
\eeal
Landau matching condition for the baryon number density requires
\beal{BaBeq}
\frac{ \Lambda_\Q^3}{ \sqrt{1+\xi_\Q}}  
\sinh\lp\frac{\lambda}{\Lambda_\Q}\rp\,{\cal H}_{\cal B}\lp \frac{m}{\Lambda_\Q},   \frac{\lambda}{\Lambda_\Q}\rp
= T^3 \sinh\lp\frac{\mu}{T}\rp\,{\cal H}_{\cal B}\lp \frac{m}{T},   \frac{\mu}{T}\rp.
\eeal
At the same time the baryon number conservation leads to the expression
\beal{BaB0}
\f{{\cal B}_{\rm 0} \tau_0}{\tau} = \frac{16 \pi k_\Q }{3} T^3\sinh\lp\frac{\mu}{T}\rp\,{\cal H}_{\cal B}\lp \frac{m}{T},   \frac{\mu}{T}\rp,
\eeal
where ${\cal B}_{\rm 0} = {\cal B}(\tau_0)$ is the initial baryon density. 

In addition to \EQS{BaBeq} and (\ref{BaB0}) we consider the following $\alpha$-weighted linear combination of Eqs.~(\ref{0mq}) so that
\begin{eqnarray}
\hspace{0cm} \f{d}{d\tau}&&\hspace{-0.75cm} 
 \left[ \alpha \Lambda_\Q^3 \left(  \tilde{{\cal H}}_{\cal N}^+\lp \f{1}{\sqrt{1+\xi_\Q}}, \frac{m}{\Lambda_\Q}, - \frac{\lambda}{\Lambda_\Q}\rp
+   \tilde{{\cal H}}_{\cal N}^+\lp \f{1}{\sqrt{1+\xi_\Q}}, \frac{m}{\Lambda_\Q}, + \frac{\lambda}{\Lambda_\Q}\rp  \right) \right. \nn \\
&& \hspace{4.25cm} 
\left. + \, (1-\alpha) \, r \,  \Lambda_\G^3 \tilde{{\cal H}}_{\cal N}^-\lp \f{1}{\sqrt{1+\xi_\G}}, 0,0\rp   \right] \nn \\
+ 
\left(\f{1}{\tau}+ \f{1}{\teq} \right) \hspace{-0.75cm} &&\left[ \alpha \Lambda_\Q^3 \left(  \tilde{{\cal H}}_{\cal N}^+\lp \f{1}{\sqrt{1+\xi_\Q}}, \frac{m}{\Lambda_\Q}, - \frac{\lambda}{\Lambda_\Q}\rp
+   \tilde{{\cal H}}_{\cal N}^+\lp \f{1}{\sqrt{1+\xi_\Q}}, \frac{m}{\Lambda_\Q}, + \frac{\lambda}{\Lambda_\Q}\rp  \right) \right. \nn \\
&& 
\hspace{4.25cm} \left. + \, (1-\alpha) \, r \,  \Lambda_\G^3 \tilde{{\cal H}}_{\cal N}^-\lp \f{1}{\sqrt{1+\xi_\G}}, 0,0\rp   \right] \nn \\
= \f{T^3}{\teq} 
&&  \hspace{-0.35cm}
\left[ \alpha \left( \tilde{{\cal H}}_{\cal N}^+\lp 1, \frac{m}{T}, - \frac{\mu}{T}\rp
+ 
 \tilde{{\cal H}}_{\cal N}^+\lp 1, \frac{m}{T}, + \frac{\mu}{T}\rp \right) + \, (1-\alpha) \, r \,  \tilde{{\cal H}}_{\cal N}^-\lp 1, 0,0\rp   \right],\nonumber\\
\label{ZM3}
\end{eqnarray}
with $r = k_\G/k_\Q = \f{4}{3}$. The functions $\tilde{{\cal H}}^\pm_{\cal N}$ are defined in Ref.~\cite{Florkowski:2017jnz}. In fact, the best agreement between the hydrodynamic results and the kinetic-theory predictions has been found  using the values $\alpha=1$ or $\alpha=0$~\cite{Florkowski:2017jnz}. The numerical results presented below are obtained with $\alpha=1$.
%
%%%%%%%%%%%%%%%%%%%%%%%%%%%%%%%%%%%%%%%%%%%%%%%%%% 
\subsection{First moments of the kinetic equations}
\label{sect:motke}
%%%%%%%%%%%%%%%%%%%%%%%%%%%%%%%%%%%%%%%%%%%%%%%%%% 
%
The first moments of \EQS{kineq} give
\beal{KEfirsthmom3}
\p_\mu  {{T}}^{\mu\nu}_{s, \rm a}  &=& U_\mu   \frac{{T}^{\mu\nu}_\eq -{T}^{\mu\nu}_{s, \rm a}}{\teq}.
\eeal
 In order to conserve energy and momentum of the entire system we require divergence of the sum of Eqs.~(\ref{KEfirsthmom3}) over ,,s'' to vanish, $\partial_\mu T^{\mu\nu}_\an=0$ (with $T^{\mu\nu}_\an = \sum_s{{T}}^{\mu\nu}_{\s, \an}$). This leads us to the conclusion that the first-moment equations~(\ref{KEfirsthmom3}) are satisfied only if the Landau matching condition for the energy density is satisfied. It can be expressed in the following way
\beal{FM1}
 &&\Lambda_\Q^4  \left( \tilde{{\cal H}}^+\lp \f{1}{\sqrt{1+\xi_\Q}}, \frac{m}{\Lambda_\Q}, - \frac{\lambda}{\Lambda_\Q}\rp
 +  \tilde{{\cal H}}^+\lp \f{1}{\sqrt{1+\xi_\Q}}, \frac{m}{\Lambda_\Q}, + \frac{\lambda}{\Lambda_\Q}\rp \right) \nn \\
 && \hspace{1cm} + \, r \, \Lambda_\G^4 \tilde{{\cal H}}^-\lp \f{1}{\sqrt{1+\xi_\G}}, 0,0\rp \nn \\
 &&   =  T^4 \left( \tilde{{\cal H}}^+\lp 1, \frac{m}{T}, - \frac{\mu}{T}\rp +  \tilde{{\cal H}}^+\lp 1, \frac{m}{T}, + \frac{\mu}{T}\rp
 + r \,  \tilde{{\cal H}}^-\lp 1, 0,0\rp   \right).
\eeal
% .
For one-dimensional  boost-invariant  systems  the  four  equations appearing in the energy-momentum conservation law, $\partial_\mu T^{\mu\nu}_\an=0$, reduce to  a single equation,
\beal{SECOND-EQUATION-1}
\f{d{\cal E}^\an  }{d\tau} = -\f{{\cal E}^\an  + {\cal P}_{L}^\an  }{\tau},
\eeal
where ${\cal P}_L^\an  = {\cal P}_L^{\Q,\an} + {\cal P}_L^{G,\an}$ is the total longitudinal momentum of the system. 
Equation~(\ref{SECOND-EQUATION-1}) can be rewritten as follows
\begin{eqnarray}
&& 
\f{d}{d\tau} \left[ \Lambda_\Q^4  \left( \tilde{{\cal H}}^+\lp \f{1}{\sqrt{1+\xi_\Q}}, \frac{m}{\Lambda_\Q}, - \frac{\lambda}{\Lambda_\Q}\rp
 +  \tilde{{\cal H}}^+\lp \f{1}{\sqrt{1+\xi_\Q}}, \frac{m}{\Lambda_\Q}, + \frac{\lambda}{\Lambda_\Q}\rp \right) \right. \nn \\
&& 
\left. \hspace{6.5cm}  + \, r \, \Lambda_\G^4 \tilde{{\cal H}}^-\lp \f{1}{\sqrt{1+\xi_\G}}, 0,0\rp \right] \nn \\ 
&=& 
-\f{1}{\tau}  \left[ \Lambda_\Q^4  \left( \tilde{{\cal H}}^+\lp \f{1}{\sqrt{1+\xi_\Q}}, \frac{m}{\Lambda_\Q}, - \frac{\lambda}{\Lambda_\Q}\rp
 +  \tilde{{\cal H}}^+\lp \f{1}{\sqrt{1+\xi_\Q}}, \frac{m}{\Lambda_\Q}, + \frac{\lambda}{\Lambda_\Q}\rp \right) \right. \nn \\
&& 
 \hspace{0.55cm} +  \Lambda_\Q^4  \left( \tilde{{\cal H}}^+_L \lp \f{1}{\sqrt{1+\xi_\Q}}, \frac{m}{\Lambda_\Q}, - \frac{\lambda}{\Lambda_\Q}\rp
 +  \tilde{{\cal H}}^+_L \lp \f{1}{\sqrt{1+\xi_\Q}}, \frac{m}{\Lambda_\Q}, + \frac{\lambda}{\Lambda_\Q}\rp \right)  \nn \\
&& 
\left. \hspace{2.5cm}  + \, r \, \Lambda_\G^4 \left( \tilde{{\cal H}}^-\lp \f{1}{\sqrt{1+\xi_\G}}, 0,0\rp
+ \tilde{{\cal H}}^-_L \lp \f{1}{\sqrt{1+\xi_\G}}, 0,0\rp \right) \right], \label{FM2} 
\end{eqnarray}
with $\tilde{{\cal H}}^\pm_L$ being introduced in Ref.~\cite{Florkowski:2017jnz}.
%%%%%%%%%%%%%%%%%%%%%%%%%%%%%%%%%%%%%%%%%%%%%
\subsection{Second moments of the kinetic equations}
\label{sect:2nd}
%%%%%%%%%%%%%%%%%%%%%%%%%%%%%%%%%%%%%%%%%%%%%
%
In order to close the system of aHydro equations we finally use the second moments of the kinetic equations \rfn{kineq}. Using Eq.~(\ref{Ioperator}), the second moments have the forms
\beal{KEsecondmom3}
\p_\lambda  {{\Theta}}_{\s, \an}^{\lambda\mu\nu}  &=& U_\lambda   \frac{{\Theta}^{\lambda\mu\nu}_{\s,\eq} -{\Theta}^{\lambda\mu\nu}_{\s, \an}}{\teq}.
\eeal
Unfortunately \EQ{KEsecondmom3} leads to an overdetermined system of equations. In order to overcome this difficulty we follow the study in Ref.~\cite{Tinti:2013vba} and select only particular combinations of equations contained in Eq.~(\ref{KEsecondmom3}), which allow us to reproduce the results of standard viscous hydrodynamics in the close-to-equilibrium limit. As a consequence we get
\begin{eqnarray}
&& \f{1}{1+\xq} \f{d\xq}{d\tau} -\frac{2}{\tau}
= - \frac{\xq  (1+\xq)^{1/2}}{\tau_{\rm eq}} \frac{T^5 }{\Lq^5} \f{\tilde{{\cal H}}_{\vartheta}^+\lp  \frac{m}{T}, - \frac{\mu}{T}\rp +\tilde{{\cal H}}_{\vartheta}^+\lp  \frac{m}{T}, + \frac{\mu}{T}\rp} {\tilde{{\cal H}}_{\vartheta}^+\lp  \frac{m}{\Lq}, - \frac{\lambda}{\Lq}\rp +\tilde{{\cal H}}_{\vartheta}^+\lp  \frac{m}{\Lq}, + \frac{\lambda}{\Lq}\rp}
, \label{sumXqgen} \\
&&  \f{1}{1+\xg} \f{d\xg}{d\tau} -\frac{2}{\tau}
= - \frac{\xg  (1+\xg)^{1/2}}{\tau_{\rm eq}} \f{T^5} {\Lg^5}.
\label{sumXggen}
\end{eqnarray} 
Equation~(\ref{sumXqgen}) refers to quarks, while~(\ref{sumXggen}) refers to gluons. The function $\tilde{{\cal H}}_{\vartheta}^+$ is defined in Ref.~\cite{Florkowski:2017ovw}.
%\%
\begin{figure}[t]
\centering
\includegraphics[width=8.0cm,clip]{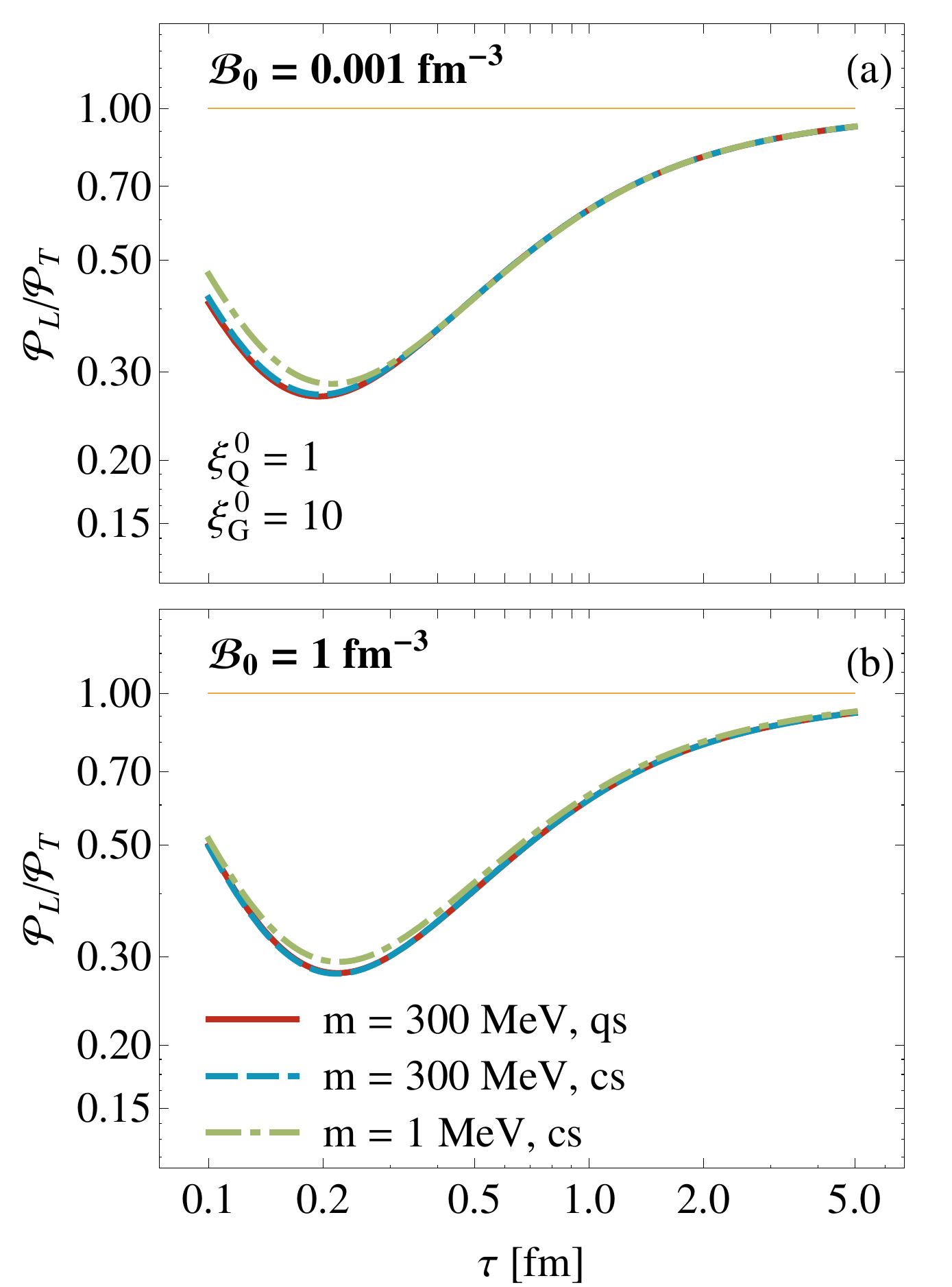}
\caption{(Color online) Time evolution of the pressure anisotropy ${\cal P}_L/{\cal P}_T$ resulting from the exact solutions of the transport equations~(\ref{kineq}). Results for the classical distributions are described by the green dashed-dotted lines ($m=1$~MeV) and blue dashed lines ($m=300$~MeV). The case of quantum statistics and $m=300$~MeV is described by the red solid lines.}
\label{fig-1}   
\end{figure}
%
%%%%%%%%%%%%%%%%%%%%%%%%%%%%%%%%%%%%%%%%%%%%%%%%%%
\section{Results}
\label{Results}
%%%%%%%%%%%%%%%%%%%%%%%%%%%%%%%%%%%%%%%%%%%%%%%%%%
%

\begin{figure}[t]
\centering
\includegraphics[width=8.0cm,clip]{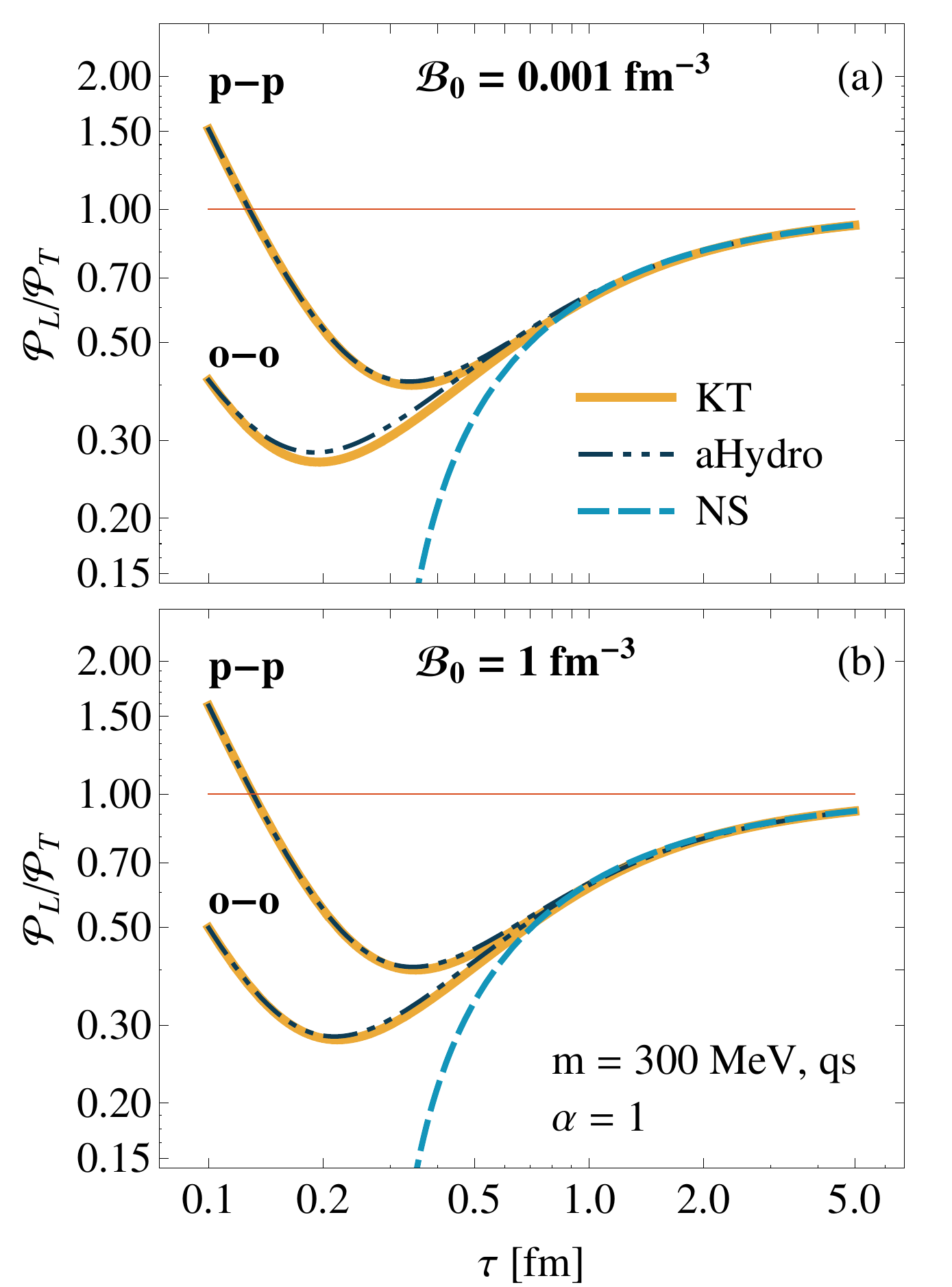}
\caption{(Color online) Exact solutions of the kinetic equations~(\ref{kineq})  (solid orange lines) compared with anisotropic hydrodynamics (black dashed-double-dotted lines) and Navier-Stokes predictions (dashed blue lines).  The results shown here were obtained  with  the parameter $\alpha$ in \EQ{ZM3} equal to unity. }
\label{fig-2}     
\end{figure}

In this section we present the kinetic theory (KT) results for  the time evolution of the pressure anisotropy ${\cal P}_L/{\cal P}_T$ of the system. Subsequently we compare them with the aHydro and Navier-Stokes (NS) predictions. Using exact solutions of the Boltzmann equation we check the influence of the finite quark mass and the quantum statistics of the constituents on the evolution of the mixture.  In the calculations we use the values $m=300$~MeV or $m=1$~MeV for the quark mass, and assume that gluons are always massless. 
Initial conditions used in the simulations correspond to two different pressure configurations, namely, the oblate-oblate configuration, where $\xi_Q^0=1$ and $\xi_G^0=10$, and prolate-prolate configuration, with $\xi_Q^0=-0.5$ and $\xi_G^0=-0.25$. The initial non-equilibrium chemical  potential, $\lambda^0$, is  chosen in such a~way that the initial baryon number density  is ${\cal B}_0=0.001\,\,{\rm fm}^{-3}$ (top panels in Figs.~\ref{fig-1} and \ref{fig-2}) or ${\cal B}_0=1\,\,{\rm fm}^{-3}$ (bottom panels in Figs.~\ref{fig-1} and \ref{fig-2}). Transverse momentum scales for quarks and gluons are equal, namely $\Lq^0=\Lg^0=1$~GeV. Relaxation time is assumed to be constant, $\teq=0.25$~fm.
%
%%%%%%%%%%%%%%%%%%%%%%%%%%%%%%%%%%%%%%%%%%%%%%%%%%
\subsection{Kinetic theory results}
\label{ssect:KinRes}
%%%%%%%%%%%%%%%%%%%%%%%%%%%%%%%%%%%%%%%%%%%%%%%%%%
%
In Fig.~\ref{fig-1} we show the proper-time dependence of the ${\cal P}_L/{\cal P}_T$ ratios obtained within the kinetic theory for three cases: (i) massless system with classical statistics (green dashed-dotted lines), (ii) the system containing massive quarks and classical statistics (blue dashed lines) and (iii) the system with massive quarks and quantum statistics (red solid lines). We assume here that the system is initially in the oblate-oblate configuration. One can observe that the ${\cal P}_L/{\cal P}_T$ ratios depend very weakly on the quark mass and the choice of statistics. At large times all curves slowly approach unity which is related to ultimate  equilibration of the system.

%%%%%%%%%%%%%%%%%%%%%%%%%%%%%%%%%%%%%%%%%%%%%%%%%%
\subsection{Anisotropic hydrodynamics results}
\label{ssect:aHydRes}
%%%%%%%%%%%%%%%%%%%%%%%%%%%%%%%%%%%%%%%%%%%%%%%%%%
%
Using aHydro equations~(\ref{BaBeq}), (\ref{BaB0}),  (\ref{ZM3}), (\ref{FM1}), (\ref{FM2}), (\ref{sumXqgen}), and (\ref{sumXggen}) we may determine various thermodynamic-like quantities, including ${\cal P}_L^\an$ and ${\cal P}_T^\an$, as functions of the proper time $\tau$. These may be compared with the exact kinetic theory results to quantify the effectiveness of the hydrodynamic approximation scheme. In  Fig.~\ref{fig-2} we present the comparison of kinetic theory results for the time dependence of the ${\cal P}_L/{\cal P}_T$ ratio with the corresponding ones obtained within the aHydro approach. We can see a very good agreement between kinetic theory and aHydro for all studied cases. In order to check the late-time behavior we present the Navier-Stokes viscous hydrodynamics predictions as well. We observe that in all cases, both aHydro and kinetic theory results approach the NS regime.

%%%%%%%%%%%%%%%%%%%%%%%%%%%%%%%%%%%%%%%%%%%%%%%%%%
\section{Summary}
\label{Summary}
%%%%%%%%%%%%%%%%%%%%%%%%%%%%%%%%%%%%%%%%%%%%%%%%%%
%
In this paper we have studied the mixture of quark, antiquark, and gluon fluids within the relativistic kinetic theory framework treated in the relaxation time approximation. We have shown that the resulting solutions are rather insensitive to the assumed quark mass value and the choice of statistics of the constituent particles. The kinetic theory solutions turn out to be in a good agreement with the corresponding results obtained within anisotropic hydrodynamics framework constructed for such a mixture. As expected, in the late-time limit, where the system approaches the close-to-equilibrium regime, all results agree very well with the Navier-Stokes approach.
%
%%%%%%%%%%%%%%%%%%%%%%%%%%%%%%%%%%%%%%%%%%%%%%%%%%
\section*{Acknowledgements}
%%%%%%%%%%%%%%%%%%%%%%%%%%%%%%%%%%%%%%%%%%%%%%%%%%
%
I would like to thank Wojciech Florkowski and Radoslaw Ryblewski for fruitful collaboration on the topics presented in this work.
%
% BibTeX or Biber users please use (the style is already called in the class, ensure that the "woc.bst" style is in your local directory)
% \bibliography{name or your bibliography database}
%
% Non-BibTeX users please use
%

\end{document}